\documentclass[3p,times,twocolumn]{elsarticle}
 \biboptions{comma,sort&compress}
 
\usepackage{graphicx}
\usepackage{amsmath}
\usepackage{dsfont}
\usepackage{here}
\usepackage{ecrc}

\volume{00}

\firstpage{1}

\journalname{Nuclear and Particle Physics Proceedings}

\runauth{}

\jid{nppp}

\jnltitlelogo{Nuclear and Particle Physics Proceedings}

\usepackage{amssymb}

\usepackage[figuresright]{rotating}

\usepackage{color}
\usepackage{ulem}

\def\m{M_{\rm QCD}}

\begin{document}

\begin{frontmatter}

\title{The Bridge Between Chiral Lagrangians and QCD Sum-Rules
 $^*$}
 \cortext[cor0]{Talk given at 22nd International Conference in Quantum Chromodynamics (QCD 19),  2--5 July 2019, Montpellier - FR}
 \author[label1]{Amir H. Fariborz}
\ead{fariboa@sunyit.edu}
\address[label1]{Department of Mathematics/Physics,  SUNY Polytechnic Institute, Utica, NY 13502, U.S.A.}
 \author[label2]{J. Ho
 }
\ead{j.ho@usask.ca}
\address[label2]{
Department of Physics and
Engineering Physics, University of Saskatchewan, Saskatoon, SK,
S7N 5E2, Canada
}
 \author[label2,label3]{A.~Pokraka
 }
\address[label3]{Department of Physics, McGill University, Montr\'eal, QC, H3A 2T8
}
\ead{andrzej.pokraka@mail.mcgill.ca}

 \author[label2]{T.G. Steele\fnref{fn1}}
   \fntext[fn1]{Speaker, Corresponding author.}
    \ead{tom.steele@usask.ca}


\pagestyle{myheadings}
\markright{ }
\begin{abstract}
Properties of the scale-factor matrices forming a bridge between the mesonic fields of chiral Lagrangians and quark-level structures of QCD sum-rules are reviewed.  
The scale-factor matrices combined with mixing angles provide a physical projection of a QCD correlation function matrix that disentangles the mesonic states.    This methodology
is illustrated for the isotriplet $a_0(980)$-$a_0(1450)$ system, and the scale factors are determined from the combined inputs of QCD sum-rules and chiral Lagrangians.
The resulting scale factors are shown to be in good agreement with the vacuum expectation values in the chiral Lagrangian framework.   The sensitivity of the scale factors on the gluon condensate QCD sum-rule parameter is explored.  
\end{abstract}

\end{frontmatter}
The interpretation of the light scalar mesons in terms of the underlying quark and gluonic substructures is one of the most challenging aspects of hadronic physics \cite{PDG,Weinberg_13,07_KZ}.  Given the highly-populated scalar states below 2 GeV, the scenario of a mixture of a two-quark nonet, a four-quark nonet, and gluonium seems natural 
\cite{Mec,close,mixing,NR04,06_F,08_tHooft,global,07_FJS4,05_FJS}.  In this scenario, the inverted scalar mass spectrum for four-quark states is an important theoretical feature that emerges in the MIT bag model \cite{Jaf} and in QCD sum-rules \cite{Zhang:2000db,Brito:2004tv,Chen:2007xr}. 

Chiral Lagrangian methods, in either the linear \cite{NR04,global,07_FJS4,05_FJS}  or non-linear models \cite{Mec,06_F,SS,BFSS2,Blk_rad,RCPT}, are founded on chiral symmetry and its breakdown, and the model parameters are determined from fits to experimental data.   For the scalar sector,  the mixings among scalar states provides implicit information on their quark substructure, with the  lighter states being primarily four-quark compositions while the heavier are primarily two-quark states \cite{BFSS2,06_F,global}. An important feature in these chiral Lagrangian analyses is the significant mixture of gluonium in two of the states \cite{close,06_F,oller}, a feature that also emerges from QCD sum-rule analyses 
(see e.g., \cite{qcdsr_glue_mix1,qcdsr_glue_mix2,qcdsr_glue_mix3,qcdsr_glue_mix4,qcdsr_glue_mix5,GSR_qq_results_mix} and review of earlier results in Ref.~\cite{narison_review}), and in other methods \cite{other_mix1,other_mix2}.

Both chiral Lagrangian and QCD sum-rule studies of scalar mesons have their limitations.  For example, although $U_A(1)$ symmetry provides a fundamental distinction between two-quark and four-quark compositions in chiral Lagrangians \cite{global,07_FJS4,05_FJS},  there is no way to distinguish between different substructures (e.g, molecular versus diquark).  This limitation can be potentially addressed by QCD sum-rules  because the field-theoretical results depend on the specific four-quark composition.  Similarly, QCD sum-rules are based on quark-hadron duality 
\cite{SVZ,Reinders:1984sr} and require parameterization of the spectral function.  In the case of broad, light, and overlapping states such as those found in the scalar sector (e.g.,  the $\sigma$ meson) a simple Breit-Wigner parameterization may be insufficient.  This limitation can potentially be overcome by guidance on resonance shapes from chiral Lagrangian approaches that have been fitted to experimental data.  A precise linkage between chiral Lagrangians and QCD sum-rules could thus lead to synergies for deeper understandings of low-energy hadronic physics and open new insights and avenues of exploration.

In Ref.~\cite{Fariborz:2015vsa} this linkage between QCD sum-rules and chiral Lagrangians was developed and validated for the $I=1$ scalar channel.  The key aspect of the resulting bridge connecting QCD sum-rules and chiral Lagrangians are scale factor matrices that combine with mixing angles to form a physical projection of the QCD correlation function onto single hadronic states.  In this proceedings article the properties of the scale factor matrices and the projection formalism will be reviewed along with the application to the $I=1$ channel.  The analysis of 
Ref.~\cite{Fariborz:2015vsa} is extended to study the dependence of the scale factors on the  gluon condensate (a key non-perturbative parameter in QCD sum-rules) to reflect its most recent
determination \cite{Narison:2018dcr}.  

The  generalized linear sigma model in the notation of Refs.~\cite{global,07_FJS4,05_FJS} contains two chiral nonets $M$ and  $M'$ respectively representing a two-quark nonet and a four-quark nonet structure.  The two nonets have the same chiral transformation properties but have different $U_A(1)$ charges: 
\begin{eqnarray}
M & \rightarrow& U_L \,  M \,  U_R^\dagger\,,  \qquad M\rightarrow e^{2i\nu}M\nonumber \\
M' & \rightarrow& U_L \,  M' \,  U_R^\dagger\,, ~\quad  M'\rightarrow e^{-4i\nu}M'~.
\label{M_trans}
\end{eqnarray}
The nonets are expressed in terms of their scalar and pseudoscalar components
\begin{eqnarray}
M & = & S + i\phi \nonumber \\
M' & = & S' + i \phi'
\label{M_SP}
\end{eqnarray}
where the two scalar meson nonets contain the two- and four-quark ``bare'' (unmixed) scalars
\begin{equation}
S=
\begin{pmatrix}
S_1^1 & a_0^+ & \kappa^+  \\
a_0^- & S_2^2 & \kappa^0 \\
\kappa^- & {\bar \kappa}^0 & S_3^3
\end{pmatrix},
\hskip 0.2cm
S' =
\begin{pmatrix}
{S'}_1^1 & {a'}_0^+ & {\kappa'}^+  \\
{a'}_0^- & {S'}_2^2 & {\kappa'}^0 \\
{\kappa'}^- & {\bar {\kappa'}}^0 & {S'}_3^3 \\
\end{pmatrix}
\label{SpMES}
\end{equation}
and similar matrices for $\phi$ and $\phi'$. 

We now map $M$ and $M'$ to their QCD counterparts
 $\m$ and $M'_{\,\rm QCD}$ having appropriate chiral transformation properties.  In particular, 
\begin{equation}
(\m)_a^b \propto ({\bar q}_R)^b ({q_L})_a  \Rightarrow \left(S_{\rm QCD}\right)_a^b = q_a {\bar q}^b
\label{SQCD}
\end{equation}
where as will be seen below there is no loss of generality in the choice of the $S_{\rm QCD}$ proportionality constant.  A similar expression exists for $S'_{\rm QCD}$  in terms of four-quark operators, but because of the many possible choices the specific form will be given below.  The QCD operators  and mesonic fields are related by scale factor matrices $I_{M}$ and $I_{M'}$ that align the mass dimensions
\begin{equation}
M=I_{M} M_{\rm QCD}\,,~M'=I_{M'} M'_{\,\rm QCD}~.
\label{M_scale}
\end{equation}
The properties of the scale factor matrices are governed by chiral symmetry.  Since both $M$ and $M_{QCD}$ have the transformation properties of \eqref{M_trans}, we have
\begin{gather}
M  \rightarrow U_LI_M M_{QCD}U_R^\dagger=I_M U_L M_{QCD} U_R^\dagger 
\nonumber\\
\Rightarrow [U_L,I_{M}]=0~.
\label{I_M_L}
\end{gather}
A similar analysis for the transformation of $M^\dagger$ yields 
\begin{equation}
 [U_R,I_{M}]=0~.
 \label{I_M_R}
\end{equation}
Eqs.~\eqref{I_M_L} and \eqref{I_M_R} must also apply to $I_{M'}$ because the chiral transformation properties of $M$ and $M'$ are identical
\begin{equation}
 [U_R,I_{M'}]=0= [U_L,I_{M'}]=0~.
 \label{I_Mp}
\end{equation}
Collectively Eqs.~\eqref{I_M_L}--\eqref{I_Mp} imply that the scale factor matrices are multiples of the identity matrix
\begin{equation}
I_M=-{m_q\over \Lambda^3}\times \mathds{1}\,,~I_{M'}={1\over {{\Lambda'}^5}}\times  \mathds{1}\, ,
\label{scale_factors}
\end{equation}
where the (constant) scale factor quantities $\Lambda$ and $\Lambda'$ have dimensions of energy  that must be determined and the quark mass factor $m_q=(m_u+m_d)/2$ has been chosen to  result in renormalization-group invariant currents as discussed below.  

The specific example of
the isotriplets $a_0(980)$ and $a_0(1450)$ illustrates the determination of the scale factors and
the detailed implementation of the bridge between chiral Lagrangians and QCD sum-rules.  The physical states are mixtures of the chiral Lagrangian fields $S$ and $S'$ which are then related to QCD operators 
\begin{eqnarray}
{\bf A}&=&
\begin{pmatrix}
a_0^0(980)\\
a_0^0(1450)
\end{pmatrix}
= L_a^{-1}
\begin{pmatrix}
\left(S_1^1 - S_2^2\right)/\sqrt{2}\\
\left({S'}_1^1 - {S'}_2^2\right)/\sqrt{2}
\end{pmatrix}
\nonumber\\
&=&
L_a^{-1} I_a J^{\rm QCD} 
\label{A_def}
\end{eqnarray}
where $L_a$ is the rotation matrix for isovectors of Ref.~ \cite{global}, $I_a$  is formed out of the scale factors defined for the two chiral nonets in (\ref{M_scale}) and $J^{QCD}$  is constructed from two- and four-quark operators
\begin{gather}
L^{-1}_a=\begin{pmatrix}
\cos\theta_a & -\sin\theta_a
\\
\sin\theta_a & \cos\theta_a
\end{pmatrix}
\,,~I_a =
\begin{pmatrix}
{-m_q\over \Lambda^3} &0 \\
0 & {1\over {{\Lambda'}^5}}
\end{pmatrix}
\\
J^{\rm QCD} =\frac{1}{\sqrt{2}}
\begin{pmatrix}
        \left( S_{\rm QCD} \right)_1^1 -
       \left( S_{\rm QCD} \right)_2^2
       \\[2pt]
\left(  S'_{\rm QCD} \right)_1^1 - \left(S'_{\rm QCD} \right)_2^2
\end{pmatrix}
\label{J_QCD}
~.
\end{gather}
Because Eq.~\eqref{A_def} defines the physical $a_0$ states, the projected physical currents $J^P$ are defined by 
\begin{equation}
J^P = L_a^{-1} I_a J^{\rm QCD}~.
\end{equation}
A physical correlation function matrix  $\Pi^P$ is then constructed from a physical projection of a QCD correlation function matrix  $\Pi^{QCD}$
\begin{gather}
 \Pi^P(Q^2) = {\widetilde {\cal T}}^a \Pi^{\rm QCD}(Q^2)  {\cal T}^a\,,~~{\cal T}^a= I_a \, L_a
 \label{phys_corr}
 \\
 \Pi^{\rm QCD}_{mn}(x) =\langle 0| {\rm T}  \left[ J^{\rm QCD}_m (x) J^{\rm QCD}_n(0) \right] |0 \rangle
\end{gather}
 where ${\widetilde {\cal T}^a}$  denotes the transpose of the matrix ${\cal T}^a$.  Although the development of \eqref{phys_corr} was in the context of the $a_0$ system, the relation for the projected physical correlator is easily generalized by the appropriate rotation matrix and scale factor matrix.
 
 The physical correlator matrix is diagonal, providing a self-consistency condition relating elements of the QCD correlation function matrix.  For example, the case of the $2\times 2$ $a_0$ system leads to the following constraint from the vanishing of off-diagonal elements\footnote{Eq.~\eqref{constraint} corrects a minor typographical error in Ref.~\cite{Fariborz:2015vsa}.}
 \begin{equation}
 \Pi_{12}^{\rm QCD} = -
\left[
{
 {    {\widetilde {\cal T}}^a_{11} \Pi_{11}^{\rm QCD} {\cal T}^a_{12}
    + {\widetilde {\cal T}}^a_{12} \Pi_{22}^{\rm QCD} {\cal T}^a_{22}
 }
 \over
 {{\widetilde {\cal T}}^a_{11}  {\cal T}^a_{22} + {\widetilde {\cal T}}^a_{12}  {\cal T}^a_{12}
 }
}
\right]~.
\label{constraint}
\end{equation}
The relation \eqref{constraint} can be used in different ways depending on whether it is feasible to calculate the off-diagonal QCD correlator.  In the $a_0$ case, the leading corrections to the off diagonal correlator are a difficult four-loop topology, so  \eqref{constraint} will be used as input for $ \Pi_{12}^{\rm QCD}$.  

In QCD sum-rule methodologies, an integral transform is applied to a dispersion relation relating the QCD  and hadronic contributions  to the projected physical correlators  \cite{SVZ,Reinders:1984sr}.  
The mixing angle matrix $L_a$ must be chosen to isolate individual states so it is important to use a sum-rule method that can check whether a residual effect of multiple states is occurring because of an insufficiently accurate mixing matrix. Laplace sum-rules are not ideal for this purpose because they suppress heavier states, so 
Gaussian sum-rules will be employed  because they provide similar weight to all states \cite{gauss,harnett_quark}.  The hadronic part of the Gaussian sum-rule is given by 
\begin{equation}
G^{\rm H} ({\hat s}, \tau) =\frac{1}{\sqrt{4\pi\tau}}
\int\limits_{s_{th}}^{\infty} \!\! dt \,{\rm exp} \left[  {{-({\hat s} - t)^2}\over {4\tau}}\right]\,\rho^H(t)~.
\label{GSR}
\end{equation}
In \eqref{GSR},  the hadronic spectral function is determined via the mesonic fields and a QCD continuum  above the  continuum threshold $s_0$.
\begin{eqnarray}
\rho^H(t)&=&
{1\over \pi} {\rm Im} \Pi^{\rm H}(t) \\
\Pi^{\rm H}_{ij} \left(q^2\right)&=&\int d^4x\, e^{iq\cdot x}
\langle 0| {\rm T} \left[ {\bf A}_i (x) {\bf A}_j(0) \right] |0 \rangle
\nonumber \\
&=&\delta_{ij} \,
\left(
{1\over {m_{ai}^2 - q^2 -i m_{ai}\Gamma_{ai}}}
+ {\rm cont.}\right)~
\label{PiH}
\end{eqnarray}
The effect of final-state interactions in the $\pi\eta$ channel is not  significant  and therefore in the first approximation is neglected here.
The last term denoted by ``cont" represents the QCD continuum contribution inherent in QCD sum-rule methods \cite{SVZ,Reinders:1984sr}.  The hadronic and QCD contributions to the Gaussians sum-rules are now equated:
\begin{equation}
G^H\left(\hat s, \tau\right)
=\widetilde{\cal T}^aG^{\rm QCD}\left(\hat s,\tau ,s_0\right){\cal T}^a
\label{full_GSR}
\end{equation}
where the QCD continuum has been absorbed from the hadronic side into the QCD contributions.  Methods for calculating the QCD prediction $G^{\rm QCD}$ from the underlying correlation function are reviewed in Ref.~\cite{harnett_quark}.  
The hadronic side of \eqref{full_GSR} is diagonal  because the states have been disentangled by the rotation matrix and the QCD side  of \eqref{full_GSR} is diagonalized by imposing the constraint \eqref{constraint}  
\begin{equation}
G^H =
\begin{pmatrix}
{(G^H)}_{11} & 0 \\
0 & {(G^H)}_{22}
\end{pmatrix}
=\widetilde{\cal T}^aG^{\rm QCD}{\cal T}^a\,.
\label{G_matrix}
\end{equation}
The resulting diagonal elements of \eqref{G_matrix} are given by 
\begin{gather}
G_{11}^H(\hat s,\tau)=a A
G_{11}^{QCD}\left(\hat s,\tau, s_0^{(1)}\right)-bB
G_{22}^{QCD}\left(\hat s,\tau, s_0^{(1)}\right)
\label{G_eqs}
\\
G_{22}^H(\hat s,\tau)=-aB
G_{11}^{QCD}\left(\hat s,\tau, s_0^{(2)}\right)+bA
G_{22}^{QCD}\left(\hat s,\tau, s_0^{(2)}\right)
\nonumber
\\
A=\frac{\cos^2\theta_a}{\cos^2\theta_a-\sin^2\theta_a}\,,~
B=\frac{\sin^2\theta_a}{\cos^2\theta_a-\sin^2\theta_a}
\\
a=\frac{m_q^2}{\Lambda^6}\,,~b=\frac{1}{\left(\Lambda'\right)^{10}}
\end{gather}
where the QCD continuum has been absorbed into the QCD Gaussian sum-rules,  $G_{11}^H$ and $G_{22}^H$ respectively represent $a_0(980)$ and  $a_0(1450)$, and the factor of $m_q^2$ is combined with  $G_{11}^{QCD}$  for renormalization-group purposes.  Note that each sum-rule has its own continuum threshold represented by $s_0^{(1)}$ and $s_0^{(2)}$, and  the constraint \eqref{constraint} has been used within the QCD prediction. 

The QCD currents in \eqref{J_QCD} used to construct the QCD sum-rules are  \cite{Chen:2007xr,harnett_quark}
\begin{gather}
J^{QCD}=\begin{pmatrix}
J_1\\
J_2
\end{pmatrix}
\,,~
J_1=\left(\bar u u-\bar d d\right)/\sqrt{2}\\
\begin{split}
J_2&=\frac{\sin\phi}{\sqrt{2}}d^T_\alpha C\gamma_\mu\gamma_5 s_\beta\left(\bar d_\alpha\gamma^\mu\gamma_5 C\bar s_\beta^T-\alpha\leftrightarrow \beta \right)
\\
&+\frac{\cos\phi}{\sqrt{2}}d^T_\alpha C\gamma_\mu s_\beta\left(\bar d_\alpha\gamma^\mu C\bar s_\beta^T+\alpha\leftrightarrow \beta \right)
- u\leftrightarrow d
\end{split}
\end{gather}
where $C$ is the charge conjugation operator and $\cot\phi=1/\sqrt{2}$ \cite{Chen:2007xr}. Expressions for the resulting QCD sum-rules can be found in Refs.~\cite{Chen:2007xr,GSR_qq_results_mix,harnett_quark} along with the necessary QCD input parameters (e.g., QCD condensates). The choice $\tau=3\,{\rm GeV^4}$ has been made, consistent with the central value used in Refs.~\cite{GSR_qq_results_mix,harnett_quark}.   The physical mass and width of the $a_0$ states are used along with 
 $\cos\theta_a=0.493$ from chiral Lagrangians \cite{global}.  Two different values of the gluon condensate are considered: $\langle\alpha G^2\rangle=0.07\,{\rm GeV^4}$ as used in 
 Ref.~\cite{Fariborz:2015vsa} and $\langle\alpha G^2\rangle=0.06\,{\rm GeV^4}$ corresponding to the lower bound determined in Ref.~\cite{Narison:2018dcr}.
 For the strange quark condensate we use $\langle \bar s s\rangle=0.8\langle\bar q q\rangle$ (see e.g., Refs.~\cite{Reinders:1984sr,Gubler:2018ctz}) in conjunction with $m_s/m_q=27.3$ \cite{PDG} and PCAC for $m_q\langle \bar q q\rangle$.
 
\begin{figure}[htb]
\centering
\includegraphics[width=\columnwidth]{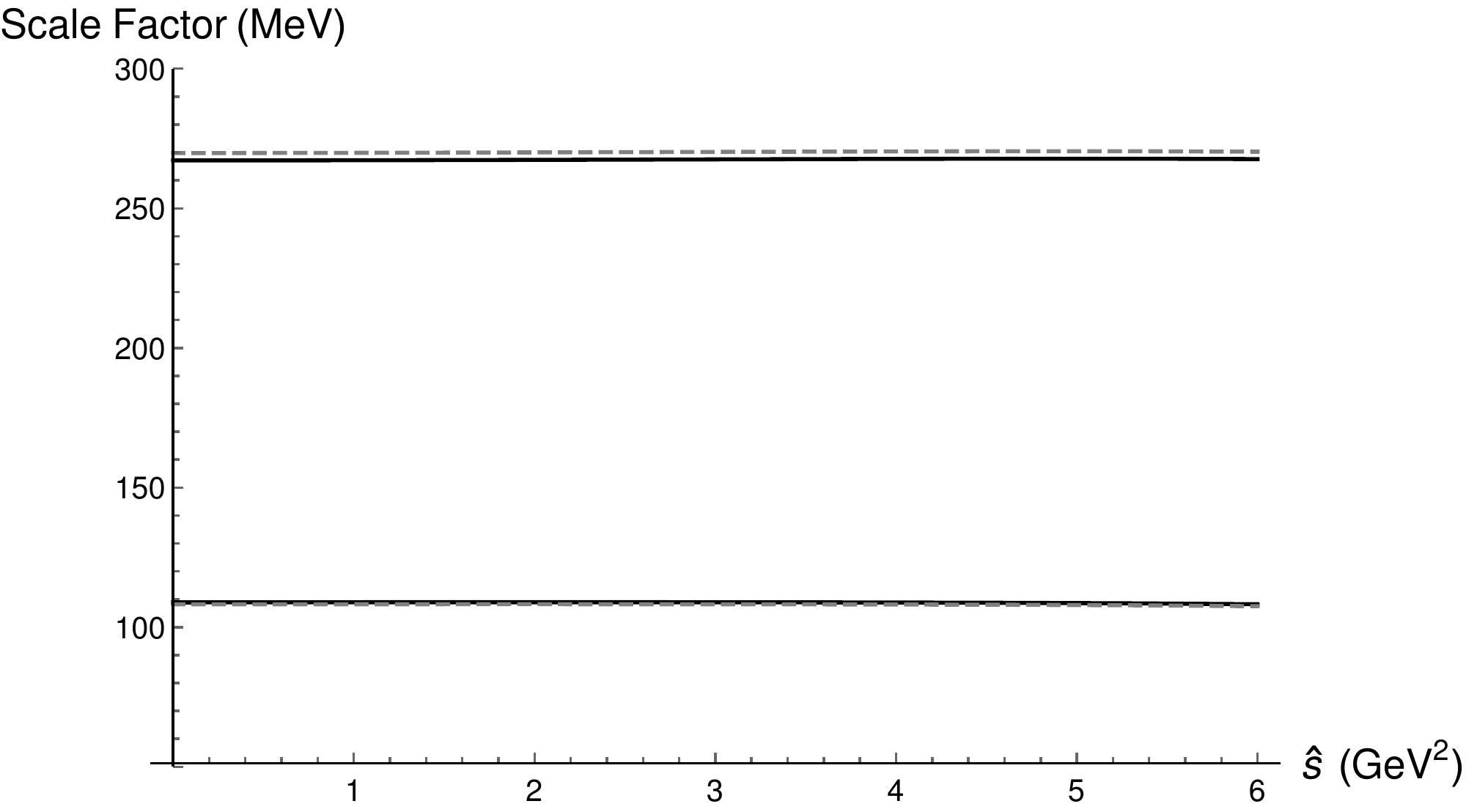}\hspace{0.02\columnwidth}
\caption{The scale factors $\Lambda$ (lower pair of curves) and $\Lambda'$ (upper pair of curves) are shown as a function of $\hat s$
for optimized continuum thresholds in Table \ref{scale_tab}.  Solid curves use the value $\langle \alpha G^2\rangle=0.06\,{\rm GeV^4}$  as QCD input  and the dashed curves are for $\langle \alpha G^2\rangle=0.07\,{\rm GeV^4}$.   The mixing angle $\cos\theta_a=0.493$ of Ref.~\cite{global} has been used along with $\tau=3\,{\rm GeV^4}$ as in Refs.~\cite{GSR_qq_results_mix,harnett_quark}.
}
\label{scale_fig1}
\end{figure}

 Eqs.~\eqref{G_eqs} are solved for the (constant) scale factors 
$\Lambda$ and $\Lambda'$, and a procedure is developed to  optimize  the continuum thresholds that minimize the $\hat s$ dependence of the scale factors.  
Fig.~\ref{scale_fig1} shows the $\hat s$ dependence of the scale factors for the optimized values of the continuum for the two values of the gluon condensate $\langle \alpha G^2\rangle$.  
The remarkable independence of the scale factors on the auxiliary sum-rule parameter $\hat s$ clearly demonstrates validity of the bridge connecting QCD sum-rules and chiral Lagrangians.  

The best-fit predictions of the scale factors and continuum thresholds are given in Table~\ref{scale_tab} along with the vacuum expectation value 
\begin{equation}
\langle S_1^1\rangle=-\frac{m_q\langle\bar u u\rangle}{\Lambda^3}\,.
\end{equation}
The QCD predictions for the vacuum expectation value are in excellent agreement with the chiral Lagrangian value $\langle S_1^1\rangle=0.056 \,{\rm GeV}$  \cite{global}.   The scale factor $\Lambda'$ 
can also be related to a vacuum expectation value using the vacuum saturation hypothesis
\begin{equation}
\langle {S'}_1^1\rangle=1.31\frac{\langle \bar d d\rangle \langle \bar s s\rangle}{\Lambda'^5}
\label{spvev}
\end{equation} 
 The QCD side of \eqref{spvev}  is not a renormalization-group invariant quantity  so it is not clear how to choose a renormalization scale for comparison.   An estimate of the QCD value yields
 $\langle {S'}_1^1\rangle\approx 0.08\,{\rm GeV}$ which is in reasonable agreement with the chiral Lagrangian value  $\langle {S'}_1^1\rangle\approx 0.03\,{\rm GeV}$ \cite{global} considering  that the  vacuum saturation hypothesis and renormalization scale dependence effects could each introduce a factor of 2--3.

\begin{table}[htb]
\begin{tabular}{rrrrrr}
\hline
\rule{0pt}{3ex}   
$\langle \alpha G^2\rangle$ &  $s_0^{(1)}$ & $s_0^{(2)}$ & $\Lambda$ & $\Lambda'$ & $\langle S_1^1\rangle$
\\[2pt]
\hline
0.07 & 1.50 & 3.14 & 0.108 & 0.270 & 0.067
\\
0.06 & 1.53 & 3.13 & 0.109 & 0.267 & 0.065
\\
\hline
\end{tabular}
\caption{Values for the optimized scale factors $\left\{\Lambda,~\Lambda'\right\}$ and continuum thresholds $\left\{s_0^{(1)},~s_0^{(2)}\right\}$ and vacuum expectation value $\langle S_1^1\rangle$ for different inputs of the gluon condensate
$\langle \alpha G^2\rangle$.  All quantities are in appropriate powers of GeV.}
\label{scale_tab}
\end{table}

\begin{figure}[htb]
\centering
\includegraphics[width=\columnwidth]{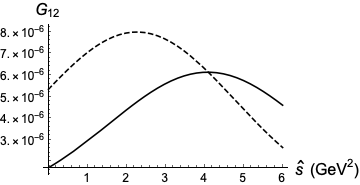}\hspace{0.02\columnwidth}
\caption{Off-diagonal Gaussian sum-rule $G_{12}$ emerging from the QCD constraint \eqref{constraint}  (solid curve) compared with the estimated leading-order perturbative QCD contribution (dashed-curve). The optimized parameters from the first row of Table~\ref{scale_tab} have been used along with $\tau=3\,{\rm GeV^4}$ as in Refs.~\cite{GSR_qq_results_mix,harnett_quark}.
All scales in GeV units.
}
\label{scale_fig2}
\end{figure}

Two other aspects were explored in the analysis.  Variations away from the chiral Lagrangian mixing angle $\cos\theta_a=0.493$ were explored, and the optimized $\chi^2$ used to measure the variations of the scale factors from a constant value increased when the mixing angle was decreased to  $\cos\theta_a=0.4$ and increased to  $\cos\theta_a=0.6$, suggesting the possibility that QCD sum-rules could help distinguish between different chiral Lagrangian mixing scenarios.  The physical-projection constraint \eqref{constraint} was also explored by comparing the Gaussian sum-rule obtained from the constraint with the estimated leading-order perturbative contribution.   The good agreement in the order-of-magnitudes emerging from  the constraint and perturbative estimates shown in Fig.~\ref{scale_fig2} implies that the physical projection of the full QCD sum-rule matrix would be nearly diagonal.   

In summary, it has been shown how chiral symmetry governs the properties of  scale factor matrices that serve as a bridge connecting QCD sum-rules and chiral Lagrangians.  
These scale factor matrices combine with a mixing matrix that projects QCD sum-rules onto physical mesonic states.   The $a_0$ system  was used to illustrate the extraction of the scale factors using the combined information from chiral Lagrangians and QCD sum-rules.  The extracted scale factors were remarkably independent of auxiliary sum-rule parameters and result in excellent agreement with renormalization-group invariant chiral Lagrangian vacuum expectation values,  providing an important validation of the methodology to bridge QCD sum-rules and chiral Lagrangians.   In future work, the connection between chiral Lagrangians and QCD sum-rules  will continue to be explored for other states in the scalar nonets.  A key test of the bridge connecting QCD sum-rules and chiral Lagrangians will be to establish the universality of the scale factors for other sectors of the the scalar nonet.  

\section*{Acknowledgments}
TGS is grateful for the hospitality of AHF and SUNY Polytechnic Institute while this work was initiated.  
TGS and JH are grateful for research funding from the Natural Sciences and Engineering Research Council of Canada (NSERC), and AHF is grateful for a 2019 Seed Grant from SUNY Polytechnic.

\end{document}